\documentclass[prd,showpacs,preprintnumbers,superscriptaddress,amsmath]{revtex4}

\usepackage{graphicx}
\usepackage{epsfig}
\usepackage{bm}

\makeatletter \leftline{\epsfbox{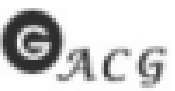}}
 \leftline
{\footnotesize{\textbf{\textsf{GACG-04/08}}}}
\begin{document}
\title{On the thermal description of the BTZ black holes}
\author{Norman Cruz}
\altaffiliation{ncruz@lauca.usach.cl}
\affiliation{Departamento de F\'\i sica, Facultad de Ciencia,
Universidad de Santiago de Chile, Casilla 307, Correo 2, Santiago,
Chile.\\}

\author{Samuel Lepe}
\altaffiliation{slepe@ucv.cl} \affiliation{Instituto de F\'\i
sica, Facultad de Ciencias B\'asicas y Matem\'aticas, Pontificia
Universidad Cat\'olica de Valpara\'\i so, Avenida Brasil 2950,
Valpara\'\i so, Chile.}

\date{\today}
\begin{abstract}
We investigate the limitations on the thermal description of three
dimensional BTZ black holes. We derive on physical grounds three
basic mass scales that are relevant to characterize these
limitations. The Planck mass in 2+1 dimensions indicate the limits
where the black hole can emit Hawking's radiation. We show that
the back reaction is meaningless for spinless BTZ black hole. For
stationary BTZ black holes the nearly extreme case is analyzed
showing that may occur a break down of its description as a
thermal object.

\end{abstract}
\maketitle

%

\section{Introduction}

An important point in discussion is if the extreme black holes
behave as thermal objects. More than a decade ago, Preskill {\it
et al} \cite{Preskill} pointed out that the thermal description
of a black hole becomes ill-defined as the black hole approaches
the extreme limit. In its treatment the use the criterion that
the standard semiclassical description, which neglects the back
reaction, is not self-consistent if the emission of a typical
quantum radiation changes the temperature by an amount comparable
to the value of the temperature. For a Schwarzschild black hole,
the thermal description breaks down simultaneously as the
classical description of spacetime does \cite{Preskill}. At the
final stages of the evaporation, the black hole field has a high
curvature and the radius of the horizon is of the order of the
Planck length, besides, as the temperature rises as the mass
diminishes, the problem of the back reaction becomes unavoidable.
Thus, the Planck mass is the only relevant scale of mass that it
is needed to take account in a full theory of quantum gravity.
Nevertheless, for a Kerr-Newman black hole the breaks down occurs
near the extreme case when the mass is much greater than the
Planck mass and the corrections due to quantum gravity are
expected to be negligible.

In $2+1$ dimensional gravity the BTZ black hole solution is a
spacetime of constant negative curvature, but it differs from
anti-de Sitter (AdS) space in its global properties \cite{BTZ}.
The related investigations of the thermodynamics properties of
this solution have not included considerations about the nature
of a possible break down of its thermal description. A obvious
question is if this type of black hole experiment, as its four
dimensional counterpart, a break down of the its thermal
description and what is the relevant mass scale involved. It was
found by Reznik \cite{Reznik} that the Planck mass in $2+1$
dimensions has a physical significance related to the description
of the thermodynamics of the Hawking emission process, for static
BTZ black holes. This is a key point when we try to shed some
light in the behavior of BTZ black holes in relation to limits to
its description as a thermal objects. The purpose of this article
is to investigate the limits in the thermal description, using
the approach describe in \cite{Preskill}, for static and rotating
BTZ black holes.

In Section II we expose briefly the principal parameters related
with the thermodynamics of the BTZ black hole. In Section III we
show the existence of three mass scales: $m_{P}$, $m_{\lambda }$,
and $m_{T}$ and discuss its physical meaning. The mass scale
$m_{T}$ has not been previously discussed in the literature; it
appears when the approach describe in \cite{Preskill} is applied
to the thermodynamics of BTZ black holes. We have restricted our
discussion to the case when the semiclassical description of the
BTZ spacetime is valid. In Section IV we discuss the particular
limitations that experiment the thermodynamical description of
BTZ black holes. We show the roll played by the above three
masses in these limitations. In particular, we explicit calculate
the conditions imposed on the black hole parameters in order to
have a well description of the thermodynamics of the Hawking
emission process. The problem of the back reaction is considered
for spinless BTZ black holes. The near extreme black hole is
discussed and the breakdown of its thermal description, in the
sense discussed in \cite{Preskill}, is analyzed. Finally, in
Section V some concluding remarks are presented.


\section{The BTZ black hole}


In this work we adopt units such that $c=1=k_{B}$, where $k_{B}$
is the Boltzmann constant. We keep the G constant explicitly in
the above expressions. $MG$ is the geometrized mass, which have no
dimensions in $2+1$ gravity. In the following section we will
discuss the mass scales relevant in the thermodynamics of the BTZ
black hole and the role of $G^{-1}$ as Planck mass. We also keep
$\hbar$ in the following thermodynamics parameters, since exist
another scale which contain explicitly this constant.

The action considered in \cite{BTZ} is
\begin{equation}
I={\frac{1}{2\pi G}}\int
\sqrt{-g}\,(R+2\ell^{-2})\,d^{\,2}x\,dt+B, \label{action}
\end{equation}
where $B$ is a surface term, and the radius of curvature
$\ell=(-\Lambda )^{-1/2}$ provides the length scale necessary to
have a horizon ($\Lambda $ is the cosmological constant).

The axially symmetric BTZ black hole written in stationary coordinates is
\begin{equation}
ds^{2}=-N^{2}\left( r\right) dt^{2}+f^{-2}\left( r\right)
dr^{2}+r^{2}(N^{\phi }\left( r\right) dt+d\phi )^{2},  \label{ds2}
\end{equation}
where the lapse $N\left( r\right) $ and the angular shift $N^{\phi }\left(
r\right) $ are given by
\begin{equation}
N^{2}\left( r\right) =f^{2}\left( r\right) =-MG+\frac{r^{2}}{\ell^{2}}+\frac{%
(JG)^{2}}{4r^{2}}\qquad \mbox{and}\qquad N^{\phi }\left( r\right) =-{\frac{JG}{%
2r^{2}}}\,,  \label{lapse}
\end{equation}
with $-\infty <t<\infty $ , $0<r<\infty $ , and $0\leq \phi <2\pi
$. The constant of integration $M$ is the conserved charge
associated with asymptotic invariance under time displacements
and $J$ (angular momentum) is that associated with rotational
invariance. The lapse function vanishes for two values of $r$
given by
\begin{equation}
\displaystyle r_{\pm }(x)=\ell (MG)^{\frac{1}{2}}
\left[\frac{1}{2}\left ( 1\pm \sqrt{x}\right) \right]
^{\frac{1}{2}}, \label{horz}
\end{equation}
where $x$ is defined by the relation
\begin{equation}
x\equiv 1-\frac{J^{2}}{(\ell M)^{2}}. \label{x}
\end{equation}
The black hole horizon is given by $r_{+}(x)$, which exist only
for $M>0$ and $|J|\leq \ell M$. We will denote the horizon of a
spinless black hole by $r_{+}(1)\equiv r_{+}(x=1)=\ell
(MG)^{\frac{1}{2}}$. The entropy of this black hole is given by
\begin{equation}
S=\left(\frac{1}{\hbar G}\right)4\pi r_{+}(x), \label{entropy}
\end{equation}
>From Eq.(\ref{entropy}) it is possible to write the entropy, $S$,
as a function of $M$ and $J$ and obtain the first law of the
thermodynamics \cite{Smarr}
\begin{equation}
dM=TdS+ \Omega dJ, \label{firstlaw}
\end{equation}
where $T$\ is the Hawking's temperature and $\Omega$ is the
angular velocity of the black hole horizon. Both parameters can
be expressed in terms of the variable $x$, yielding for the
temperature
\begin{equation}
T\left( x\right)=T\left( 1\right)
\sqrt{\frac{2x}{1+\sqrt{x}}},\label{temperx}
\end{equation}
where $T(1)$ is the temperature of an spinless BTZ black hole
given by
\begin{equation}
T\left( 1\right) =\frac{\hbar G
}{2\pi}\frac{r_{+}(1)}{\ell^{2}}.\label{temperstatic}
\end{equation}
The corresponding expression for $\Omega$ is
\begin{equation}
\Omega=\frac{1}{\ell}\frac{\sqrt{1-x}}{1+\sqrt{x}}
\end{equation}
Important aspects of the thermodynamics systems need to know the
heat capacities $C_{z}$, where $z$ denote the set of parameters
held constant. A useful expression, that we will use latter, for
the heat capacity at constant angular momentum $J$ is given by
\begin{equation}
C_{J}=T\left( \frac{\partial S}{\partial T}\right)_{J}=
C_{0}\,\frac{1}{2-\sqrt{x}}\left[\frac{(1+\sqrt{x})x}{2}\right]^{1/2}
, \label{capacityX}
\end{equation}
where $C_{0}=\left( \hbar G\right) ^{-1}4\pi r_{+}$ is the heat
capacity at $J=0$. The expression for $C_{\Omega}$ is given by
\begin{equation}
C_{\Omega}= T\left( \frac{\partial S}{\partial
T}\right)_{\Omega}= \frac{4\pi \ell}{\hbar G}\left(\frac{MG}{2}
\left [1+\sqrt{x}\right]\right)^{1/2}. \label{Comega}
\end{equation}
Is direct to verify that $C_{J}$ and $C_{\Omega}$ are always
positive, independently of the values of the parameters $M$ and
$J$, for the BTZ black holes. These type of black holes never
experiment a phase transition. In the case of two independent
thermodynamics variables $Z_{1}$, $Z_{2}$, stability requires
that $\partial_{Z_{1}}\partial_{Z_{1}}S\leq 0$,
$\partial_{Z_{2}}\partial_{Z_{2}}S\leq 0$,
$(\partial_{Z_{1}}\partial_{Z_{2}})^{2}S-
(\partial_{Z_{1}}\partial_{Z_{1}})S
(\partial_{Z_{2}}\partial_{Z_{2}})S\leq 0$ \cite{Callen}. It is
straightforward to prove that the above conditions are satisfied
taking  $Z_{1}= M$, $Z_{2}=J$, for static and rotating BTZ black
holes, which ensured the thermal stability of this type of black
holes. These conditions are always satisfied for any $M$ and $J$.


\section{Mass scales}


\bigskip We will show in this section that the thermal description
of a spinless BTZ black hole is ill defined when the black hole
mass approaches to a new mass scale, not discussed previously in
the literature. For this reason we first discuss the relevant mass
scales in the theory of black hole in $2+1$ dimensions.

For a Schwarszchild black hole, at the Planck scale, the
fluctuations of the geometry become important and this occurs
when its radius becomes comparable to the Compton wavelength,
{\it i.e.}, $r_{horizon}\sim \lambda \,_{Compton}$. From this
relationship we obtain the expression for the Planck mass,
$m_{P}$. Nevertheless, we can also obtain the Planck mass
imposing that $r_{horizon}\sim \ell_{P}$, where $\ell_{P}$ is the
Planck length. In addition, we can define $m_{P}$ as $\hbar
/\ell_{P}$ from a straightforward dimensional analysis. In four
dimensions, these criteria leads to a unique mass scale, the
Planck mass, given by $m_{P\,\,3+1}=(\hbar/ G)^{1/2}$.

The situation is quite different in three dimensional gravity. In general,
for any dimension, the Planck mass, $m_{P}$, is defined from the relation $%
m_{P}\ell_{P}\sim \hbar $, where $\ell_{P}$ is fundamental length
scale which is obtained imposing that the action in $D$ dimensions
\begin{equation}
I\sim \frac{1}{G}\int d^{D}x\sqrt{-g}R,  \label{actionD}
\end{equation}
be of the order of $\hbar $. For $2+1$ dimensions the fundamental unit of
length is given by
\begin{equation}
\ell_{P}=\hbar G ,  \label{LP}
\end{equation}
which leads to the corresponding Planck mass in $2+1$ dimensions
\begin{equation}
m_{P}=\frac{1}{G}  \label{MP}.
\end{equation}

We restrict our discussion to the case when the semi-classical
description of the BTZ spacetime is valid, {\it i.e.}, when the
condition $S_{2+1}/\hbar >1$ holds or, in other words, when
$\ell>\ell_{P}$.

Since the Planck mass in $2+1$ dimensions is a classical mass
unit (contains only the gravitational constant), it is not
surprising that appears in classical phenomena associated with
$2+1$ dimensional fluids in hydrostatic equilibrium. If the
cosmological constant is not included, classical results show
that there exist a universal mass, in the sense that all
rotationally invariant structures in hydrostatic equilibrium have
a mass that is proportional to $m_{P}$ \cite{Cornish}. In this
case there are no black hole solution and the possibility of
collapse is clearly forbidden. Nevertheless, the study of the
structures, with a mass $M$ and a radius $R,$ in hydrostatic
equilibrium in AdS gravity leads to an upper bound on the ratio
$M/R$ similar to the four dimensional case. This result shows
that exist the possibility of collapse for matter distributions
that have the ratio $M/R$ over the above upper bound. Black holes
with large masses are possible to exist (large with respect to
$m_{P}$), even more, any fluid
distribution in hydrostatic equilibrium has necessarily a mass greater than $%
m_{P}$. With the assumption that the BTZ black hole is the end of
a collapse, the classical mass $m_{P}$ represents the lowest mass
of any $2+1$ black hole \cite{Cruz}.

For the Planck mass, $m_{P}$, $r_{+}\sim \ell$, {\it i.e.}, the
size of the horizon is comparable with the associated length of
the space-time curvature $l$. Since $\ell >\ell_{P}$, the
fluctuations of the black hole geometry are not important at this
mass scale.

A remarkable result, as we shall see below, is that without the
length scale provides by the cosmological constant is not
possible to build a unit mass containing $\hbar$. Quantum
phenomena are presents in $2+1$ dimensions with a length scale
that can provide horizons and the basic mass units related
contain both, the Planck and the cosmological constant.

The fluctuations of the black hole geometry become important when
$r_{+}\sim \lambda _{Compton}$, {\it i.e.}, when the radius of the
black hole becomes comparable to the Compton wavelength. This
yield the mass scale, $m_{\lambda }$, given by
\begin{equation}
m_{\lambda }=\left( \frac{\hbar ^{2}}{\ell^{2}G}\right) ^{1/3}.
\end{equation}
For this reason Reznik \cite{Reznik} identify the Planck mass in
$2+1$ dimensions with $ m_{\lambda }$.

We can obtain another mass scale when the limitations of the
thermal description of a black hole are studied. In the following
analysis we will consider a classical background geometry,
ignoring the back reaction. As it was pointed out in
\cite{Preskill} the semiclassical description of the black hole
evaporation is not self consistent if the emission of a typical
quantum radiation changes the temperature by an amount comparable
to the value of the temperature. If $T$ is the energy of the
quantum and if $\Delta T$ is the change of temperature that
experiment the black hole after the emission, then from the
relation $C_{J}\Delta T=T$, where $C_{J}$ is specific heat at
$J=const.$, the condition for the thermal description to be
self-consistent is
\begin{equation}
|T \left( \frac{\partial T}{\partial (MG)}\right) _J| << |T|,
\label{termica}
\end{equation}
which is equivalent to impose
\begin{equation}
\frac{\partial T}{\partial (MG)}=C_{J}^{-1}=C^{-1}_{0} \left (
2-\sqrt{x}\right )\left[\frac{2}{(1+\sqrt{x})x}\right]^{1/2}<< 1.
\label{Wilchek}
\end{equation}
For the particular case of a static BTZ black hole, $x=1$, the
heat capacity given by
\begin{equation}
C_{0}=4 \pi \sqrt{\frac{M}{m_{T}}}, \label{Cacstatic}
\end{equation}
where mass scale, $m_{T}$, has the following expression
\begin{equation}
m_{T}=\frac{\hbar ^{2}G}{\ell^{2}}.
\end{equation}
The breakdown occurs when the black hole mass $M$ satisfy $M\sim
m_{T}$. At this scale $r_{horizon}\sim \ell_{P}\ $ and
corrections due to quantum gravity are expected to be very
important. Note that this mass satisfy (up to a numerical
factor), $T(M=m_{T})\sim m_{T}$. For Schwarzschild black holes
the corresponding relation is $T(M=m_{P\,\,3+1})\sim
m_{3+1\,\,P}$; {\it i.e.}, black holes with masses of the order
of the Planck mass radiates at Planck temperature.


\section{Limitations on the thermal description}


\subsection{The static black hole}

A crucial point of the characteristic behavior of thermodynamics
of the spinless BTZ black hole was pointed out by Reznik
\cite{Reznik}. He indicated that the physical significance of the
mass unit $m_{P}$ is that for $M>m_{P}$ ($M<m_{P}$) the
wavelength $\lambda$ of the Hawking radiation satisfies
$\lambda<r_{+}$ ($\lambda>r_{+}$).

Since the process of energy emission can be thermodynamically
well described when a typical wavelength of the Hawking radiation
satisfy $\lambda\lesssim r_{+}$, we calculate for rotating BTZ
black hole the conditions on the parameters $M$ and $J$ in order
to satisfy this restriction. We have the following relation for a
typical wavelength, $\lambda$, related to Hawking's temperature
\begin{equation}
T(x)\ell_{P}^{-1}\sim \lambda^{-1}, \label{TsimLamb}
\end{equation}
The requirement for the wavelength is
\begin{equation}
\lambda \lesssim r_{+}(x), \label{Lambhoriz}
\end{equation}
>From the relations (\ref{TsimLamb}) and (\ref{Lambhoriz}) we
obtain that
\begin{equation}
\left( \frac{r_{+}(1)}{\ell}\right)^{2}\gtrsim
\frac{1}{\sqrt{x}}. \label{horizx}
\end{equation}
>From this inequality we obtain, in terms of the dimensionless
parameters $j \equiv JG/\ell$ and $m\equiv M/m_{P}$, that
\begin{equation}
j \lesssim m\left( 1- \frac{1}{m^{4}}\right)^{1/2}. \label{jandm}
\end{equation}
Notice that for a static BTZ black hole $x=1$ and the inequality
(\ref{horizx}) yields $M\gtrsim m_{P}$, since
$r_{+}(1)/\ell=M/m_{P}$. This was the result obtained in
\cite{Reznik}.

Let us discuss first the case of a spinless BTZ black hole. Since
we only consider the regime where $\ell>\ell_{P}$ the mass scales
obtained on physical grounds satisfy the relation
$m_{T}<m_{\lambda}<m_{P}$. We have argue that for this black hole
the Hawking's radiation cannot be emitted if $M<m_{p}$. Notice
that this is an effect due to the finite size of our system, which
has no equivalent in the case of a Schwarzschild or Kerr black
holes. The other mass scales founded allow to characterize
further the limitations in the description of the thermodynamical
behavior of the BTZ black hole.

At the scale of $m_{P}$ and unlike the Schwarzschild black hole,
where the curvature of the spacetime is associated with the size
of the black hole horizon, the BTZ black hole is an AdS space of
constant curvature, which no changes through all the evaporation
process, neglecting the back reaction, with a radius of curvature
given by $\ell$. On the other hand, $r_{+}(1)> \lambda
_{Compton}$ ($m_{\lambda}<m_{P}$) which means that the
fluctuations of the black hole geometry never becomes important
for the spinless BTZ black hole.

Contrary to the Schwarzschild black hole, at the scale of $m_{P}$
the temperature obey the relation $T(M=m_{P})< m_{P}$. At this
scale the energy emitted by the black hole is no important with
respect to the energy of the black hole itself. This is
consistent with the fact that the thermal description is well
defined only for black hole masses greater than $m_{T}$
($m_{T}<m_{P}$). Or in other words, there is no breakdown of the
thermal description, in the sense discussed in \cite{Preskill},
for the spinless BTZ black hole.

Notice that the fluctuations, both in temperature and entropy,
are important at mass scale $m_{T}$. The equilibrium
thermodynamics fluctuations of BTZ black holes in the
microcanonical ensemble, canonical ensemble and grand canonical
ensemble was studied in \cite{Cai}. The fluctuations in the
temperature (microcanonical ensemble) are given by
\begin{equation}
\frac{\langle \delta T(1)\delta
T(1)\rangle}{T^{2}(1)}=C_{0}^{-1}=\frac{1}{4\pi}\sqrt{\frac{m_{T}}{M}},
\end{equation}
and in the entropy (canonical ensemble) by
\begin{equation}
\langle \delta S\delta S\rangle=C_{0}=4\pi\sqrt{\frac{M}{m_{T}}}.
\end{equation}
The above equations allow us to say that at the Planck scale, the
fluctuations of the temperature are no important and that the
system have many states sampled spontaneously.

These results have important consequences for the back reaction
problem. Mart\'{i}nez and Zanelli \cite{Martinez} have evaluated
the back reaction of a massless conformal scalar field on the
geometry of the spinless BTZ black hole. The authors calculate
the $\textit{O}$($\hbar r^{-1}$) corrections to the metric, which
do not change the value of the ADM mass. The corrections to the
temperature and entropy are linear in a function $F(M)\sim
e^{-\pi M/m_{P}}$, which implies that the back reaction becomes
large only for small masses compared to the Planck mass. These
results are consistent with the fact that at the mass scales lower
than the Planck mass, for example, $m_{T}$, the fluctuations of
the temperature are important. Nevertheless, since only has
physical sense to consider black holes above the Planck scale
calculations relative to the back reaction are meaningless.

For $\ell=\ell_{P}$, there is a unique mass scale, since
$m_{P}\sim m_{\lambda }\sim m_{T}$. In this case is no longer
valid the semiclassical description of the AdS spacetime due to
the fluctuations of the geometry becomes important, independent
of the black hole radius. It has no physical sense to consider
the limitations on the thermal description.




\subsection{The rotating black hole}

Is direct to see that the rotating BTZ black hole has a very
different behavior with respect to its four dimensional
counterparts. Rewritten Eq.(\ref{horizx}) in terms of the black
hole mass we obtain
\begin{equation}
M \gtrsim m_{P}x^{-1/4}. \label{massx}
\end{equation}
For a rotating black hole $x<1$, which implies that the allowed
minimum mass is always greater than the Planck mass. On the other
hand, if, for example, $M \approx m_{P}$, Eq. (\ref{jandm})
imposes an upper bound on the angular momentum of the black hole,
which in this case indicates that $j\ll m$. In simple words, for
black hole with a mass near to the Planck mass the upper bound on
the angular momentum is very law.

As the black hole mass increases the upper bound on the angular
momentum allowed also increases. For a massive rotating black
hole, i.e., $m\gg 1$, or equivalently, $M\gg m_{P}$, we obtain
that $j\lesssim m$. This is the only type of rotating BTZ black
hole that can approach to the extreme case. To investigate the
breakdown of the thermal description, in the sense discussed in
\cite{Preskill}, we approach to the the extreme case, $|J|\leq
\ell M$, taking $x \simeq 0$ in Eq. (\ref{Wilchek}), which yields
\begin{equation}
\frac{\partial T}{\partial (MG)}\simeq \frac{1}{2\pi}\left(
\frac{2m_{T}}{Mx}\right)^{1/2}<<1. \label{dtadm}
\end{equation}
Notice that for near extreme BTZ black holes, the thermal
description breaks down when $Mx\sim m_{T}$, which implies that
this breakdown may occur within the regime $M\gg m_{T}$. In this
case it is straightforward to see that with an adequate ratio
$\ell/ \ell_{P}$ it is possible that a massive rotating black hole
experiment a breakdown of its thermal description. The
temperature fluctuations (microcanonical ensemble) can be
evaluated from the relation
\begin{equation}
\frac{\langle \delta T(x)\delta T(x)\rangle}{T^{2}(x)}=C_{J}^{-1},
\end{equation}
which according to the rhs. of Eq. (\ref{dtadm}) implies that as
the near extreme case is approached, the temperature fluctuations
become larger to the temperature itself. The entropy fluctuations
(canonical ensemble) are given by
\begin{equation}
\langle \delta S\delta S\rangle=C_{J}\simeq 2\pi\left(
\frac{Mx}{2m_{T}}\right)^{1/2}. \label{deltaS}
\end{equation}
So near the extreme case few states are sampled spontaneously.
Notice the above arguments are valid if $Mx\rightarrow 0$ as the
the black hole mass increases, which is correct since Eq.
(\ref{jandm}) implies that $x\sim m^{-4}$ for massive black holes.

In this scenario, as we explained above, the constant curvature
of the AdS spacetime at the horizon is small in Planck units, and
corrections due to quantum gravity are expected to be negligible.
For extreme Kerr-Newman black holes it was found in
\cite{Preskill} that thermal description breaks down when $M\gg
m_{P}$, so the situation is equivalent in the sense that we are
in a scenario in which it is no necessary to include quantum
gravity.


\section{Conclusions}


We have shown the existence of three mass scales: $m_{P}$,
$m_{\lambda }$, and $m_{T}$, which have a clear physical meaning.
We have restricted our discussion to the case when the
semiclassical description of the BTZ spacetime is valid, {\it
i.e.}, when the condition $S_{2+1}/\hbar >1$ holds or, in other
words, when $\ell>\ell_{P}$. In this case the masses satisfy the
relation $m_{T}<m_{\lambda }<m_{P}$.

We have argued that the process of energy emission can be
thermodynamically well described when a typical wavelength of the
Hawking radiation satisfy $\lambda\lesssim r_{+}$.

We have shown that the radiation process take place only for
static black holes with masses above the Planck scale.

At this stage the size of the horizon is comparable with the
associated length of the spacetime curvature and the fluctuations
of the black hole geometry are not important. This means that in
the evaporation of the BTZ black hole the horizon never lost its
classical meaning. On the other hand, we have proved that the
breakdown of the thermal description, in the sense discussed by
Preskill {\it et al} \cite{Preskill}, never occurs. This is
consistent with the result found in \cite{Martinez}, in which the
back reaction becomes large for small masses compared to the
Planck mass. In this sense, the BTZ black hole never experiment a
back reaction comparatively important with respect to the
background geometry.

We have shown that for rotating BTZ black hole the allowed
minimum mass is always greater than the Planck mass. Exist,
besides, an upper bound on the angular momentum of the black hole
which depends on the black hole mass. Only for massive rotating
black holes it is possible $j\lesssim m$, and the near extreme
case can be reached. In this case the description of black holes
as thermal objects may experiment a break down. This also occurs
for Kerr-Newman black holes in $3+1$ dimensions \cite{Preskill}.


\section*{Acknowledgements}

Useful discussions with members of the GACG are gratefully
acknowledged. S. Lepe acknowledges Departamento de F\'\i sica,
Universidad de Santiago de Chile, for hospitality. The authors
also acknowledged the Universidad de Concepci\'on for their kind
hospitality. This work was supported by CONICYT through Grant
FONDECYT $N^{0}$ 1040229, USACH-DICYT $N^{0}$ 04-0031CM (NC) and
PUCV-DI $N^{0}$ 123.769/03 (SL). The authors were partially
supported by the Ministerio de Educaci\'on through a MECESUP
Grant, USA 0108. This work was partially realized during one of
the Dichato Cosmological Meetings, Concepci\'on, Chile (2004).

\end{document}